\documentclass[%
 aip,
cp,  
 amsmath,amssymb,
 reprint,%
]{revtex4-2}

\usepackage{graphicx}
\usepackage{dcolumn}
\usepackage{bm}

\usepackage[utf8]{inputenc}
\usepackage[T1]{fontenc}
\usepackage{mathptmx} 

\usepackage{amsthm}
\newtheorem{thm}{Theorem}
\newtheorem*{cor}{Corollary}
\usepackage{wasysym}
\usepackage{color}
\usepackage{hyperref}

\begin{document}

\title{The Scale Invariant Vacuum Paradigm:\\Main Results plus the Current BBNS Progress}

\newcommand{\orcidauthorA}{0000-0001-8744-0444} 
\newcommand{\orcidauthorV}{0000-0002-2022-6432} 

\author{Vesselin G. Gueorguiev}
\email[Corresponding author: ]{Vesselin\;at\;MailAPS.org}
\affiliation{Institute for Advanced Physical Studies, 1784 Sofia, Bulgaria,}
\affiliation{Ronin Institute for Independent Scholarship, 127 Haddon Pl., Montclair, NJ 07043, USA.}

\author{Andre Maeder}
\email{Andre.Maeder\;at\;UniGe.ch}
\affiliation{Geneva Observatory, University of Geneva, Chemin des Maillettes 51, CH-1290 Sauverny, Switzerland.}

\begin{abstract}
We summarize the main results within the Scale Invariant Vacuum (SIV) paradigm  
as related to the Weyl Integrable Geometry 
(WIG) as an extension to the standard Einstein General Relativity (EGR).
After a short sketch of the mathematical framework,
the main results until 2023 \cite{SIVProgress22} 
are highlighted in relation to: 
the inflation within the SIV \citep{SIV-Inflation'21},  
the growth of the density fluctuations \citep{MaedGueor19}, 
the application of the SIV to scale-invariant dynamics of galaxies, MOND, 
dark matter, and the dwarf spheroidals \citep{MaedGueor20b},
and the most recent results on the BBNS light-elements' abundances within the SIV \citep{VG&AM'23}.
\end{abstract}


\maketitle


\section{Motivation}

The paper is a summary of the current main results within the
Scale Invariant Vacuum (SIV) paradigm as related to the Weyl Integrable Geometry 
(WIG) as an extension to the standard Einstein General Relativity (EGR) as of Summer 2023.
As such, it is a reflection of the corresponding online conference presentation during the 
XXIII International Meeting Physical Interpretations of Relativity Theory at the
Bauman Moscow State Technical University, Moscow, 2023.

After a general introduction on the problem of scale invariance and physical reality,
along with the similarities and differences of Einstein General Relativity and 
Weyl Integrable Geometry, we only highlight the mathematical framework as pertained to 
Weyl Integrable Geometry, Dirac Co-Calculus, and reparametrization invariance|.
Rather than re-deriving the weak-field SIV results for the equations of motion,
we have decided to use the idea of reparametrization invariance \citep{sym13030379} 
to illustrate the corresponding  equations of motion. The relevant discussion 
on reparametrization invariance is in the section on the 
Consequences of Going beyond Einstein's General Relativity. 
This section precedes the brief review of the necessary results 
about the Scale Invariant Cosmology idea needed in the section on Comparisons and Applications, 
where we highlight the main results related to inflation within the SIV \citep{SIV-Inflation'21}, 
the growth of the density fluctuations \citep{MaedGueor19}, 
and the application of the SIV to scale-invariant dynamics of galaxies, MOND, 
dark matter, and the dwarf spheroidals \citep{MaedGueor20b}. 
The results section of the paper concludes with the most recent results on the BBNS light-elements' abundances within the SIV \citep{VG&AM'23}.
We end the paper with a section containing the Conclusions and Outlook for future research directions.

\subsection{Scale Invariance and Physical Reality}

The presence of a scale is related to the existence of physical connection and causality.
The corresponding relationships are formulated as physical laws dressed in mathematical expressions.
The laws of physics ({numerical factors in the} formulae) change upon change of scale, 
{but maintain a form-invariance}. As a result, using consistent units is paramount in physics and leads to powerful dimensional estimates
of the order of magnitude of physical quantities {based on a simple dimensional analysis}.
The underlined scale is closely related to the presence of material content,
{which reflects the energy scale involved}.

However, in the absence of matter, a scale is not easy to define. 
Therefore, an empty universe would be expected to be scale invariant!
Absence of scale is confirmed by the scale invariance of the 
Maxwell equations in vacuum (no charges and no currents---the sources of the electromagnetic fields).
The field equations of general relativity are scale invariant for empty space with zero cosmological constant.
What amount of matter is sufficient to kill scale invariance is still an open question.
Such a question is particularly relevant to cosmology and the evolution of the universe.

\subsection{Einstein {General Relativity} (EGR) and Weyl Integrable Geometry (WIG)}

Einstein's General Relativity (EGR) is based on the premise of a torsion-free covariant connection 
that is metric-compatible and guarantees the preservation of the length of vectors along geodesics
$(\delta\left\Vert \overrightarrow{v}\right\Vert =0)$.
The theory has been successfully tested at various scales, starting from local Earth laboratories,
the Solar system, on galactic scales via light-bending effects, 
and even on an extragalactic level via the observation of gravitational waves.
The EGR is also the foundation for modern cosmology and astrophysics. 
However, at galactic and cosmic scales, some new and mysterious 
phenomena have appeared. The explanations for these phenomena are often 
attributed to unknown matter particles or fields that are yet to be detected 
in our laboratories---dark matter and dark energy.

As no new particles or fields have been detected in the Earth labs
for more than twenty years, it seems reasonable to revisit some old ideas 
that have been proposed as a modification of the EGR. 
In 1918, Weyl proposed and extension by adding local gauge (scale) invariance \cite{Weyl23}.
Other approaches were more radical by adding extra dimensions, 
such as Kaluza–Klein unification theory. Then, via Jordan conformal equivalence,
one comes back to the usual 4D spacetime as projective relativity theory, but with
at least one additional scalar field. Such theories are also known as Jordan–Brans–Dicke  
scalar-tensor gravitation theories \cite{Brans14,Faraoni+99}. 
In most such theories, there is a major drawback---a varying Newton constant $G$. 
As no such variations have been observed, we prefer to view Newton's gravitational 
constant $G$ as constant despite the experimental issues on its measurements \cite{Xue20}.

In the light of the above discussion one may naturally ask: could 
the mysterious (dark) phenomena be artifacts of non-zero
$\delta\left\Vert \overrightarrow{v}\right\Vert $, 
but often negligible; thus, almost zero value $(\delta\left\Vert \overrightarrow{v}\right\Vert \approx0)$,
which could accumulate over cosmic distances and fool us that the 
observed phenomena may be due to dark matter and/or dark energy?
An idea of extension of EGR was proposed by Weyl as soon as 
{the General Relativity (GR)} was proposed by Einstein. 
Weyl proposed an extension to GR by adding local gauge (scale) invariance 
that has the consequence that lengths may not be preserved upon parallel transport.
However, it was quickly argued that such a model will result in a path dependent phenomenon 
and, thus, contradicting observations. A remedy was later found to this objection 
by introducing Weyl Integrable Geometry (WIG), where the lengths of vectors  
are conserved only along closed paths ($\varoint\delta\left\Vert \overrightarrow{v}\right\Vert =0$).
Such formulation of the Weyl's original idea defeats the Einstein objection!
Furthermore, given that all we observe about the distant universe are waves
that reach us, the condition for Weyl Integrable Geometry is basically saying 
that the information that arrives to us via different paths is interfering constructively 
to build a consistent picture of the source. 

One way to build a WIG model is to consider conformal transformation 
of the metric field $g'_{\mu\nu}=\lambda^{2}g_{\mu\nu}$ and to apply it
to various observational phenomena. As shown previously \cite{SIVProgress22},
the demand for homogeneous and isotropic space restricts the field $\lambda$ 
to depend only on the cosmic time and not on the space coordinates.
The weak field limit of such a WIG model results in an extra acceleration in the
equation of motion that is proportional to the velocity.
This behavior is somewhat similar to the Jordan–Brans–Dicke  scalar-tensor gravitation;
however, the conformal factor $\lambda$ does not seems to be a typical 
scalar field as in the Jordan–Brans–Dicke theory \cite{Brans14,Faraoni+99}.
The Scale Invariant Vacuum (SIV) idea provides a way of finding out the
specific functional form of $\lambda(t)$ as applicable to LFRW cosmology
and its WIG extension \cite{SIVProgress22, Maeder17a}. 

We also find it important to point out that extra acceleration in the
equations of motion, which is proportional to the velocity of a particle, 
could also be justified by requiring re-parametrization symmetry.
Not implementing re-parametrization invariance in a model could lead
to un-proper time parametrization \cite{sym13030379} 
that seems to induce ``fictitious forces'' in the equations of motion 
similar to the forces derived in the weak field SIV regime. It is a 
puzzling observation that may help us understand nature better.

\section{Mathematical Framework}
The  framework for the Scale Invariant Vacuum paradigm is based on the 
Weyl Integrable Geometry and Dirac co-calculus as mathematical tools for 
description of \mbox{nature \cite{Weyl23,Dirac73}}. 

The original Weyl Geometry uses a metric tensor field $g_{\mu\nu}$, 
along with a ``connexion'' vector field $\kappa_{\mu}$, and a scalar field $\lambda$.
In the Weyl Integrable Geometry, the ``connexion'' vector field $\kappa_{\mu}$ is not an independent, 
but it is derivable from the scalar field $\lambda$ via the defining expression:
$\kappa_{\mu}=-\partial_{\mu}\ln(\lambda)$.
This form of the ``connexion'' vector field $\kappa_{\mu}$ guarantees its irrelevance, 
in the covariant derivatives, upon integration over closed paths.
That is, $\varoint \kappa_{\mu}dx^{\mu} =0$. In other words, $\kappa_{\mu}dx^{\mu}$ represents a closed 1-form; 
furthermore, it is an exact form since its definition implies $\kappa_{\mu}dx^{\mu}=-d\ln{\lambda}$.
Thus, the scalar function $\lambda$ plays a key role in the Weyl Integrable Geometry.
Its physical meaning is related to the freedom of a local scale gauge, which provides 
a description upon scale change via local re-scaling $l'\rightarrow\lambda(x)l$.

The covariant derivatives use the rules of the Dirac co-calculus  \cite{Dirac73} 
where tensors also have co-tensor powers based on the way they transform upon change of scale.
For the metric tensor $g_{\mu\nu}$ this power is $n=2$. 
This follows from  the way the length of a line segment $ds$ 
with coordinates $dx^{\mu}$ is defined via the usual expression $ds^2=g_{\mu\nu}dx^{\mu}dx^{\nu}$. 
That is, one has:
$l'\rightarrow\lambda(x)l\Leftrightarrow ds'=\lambda ds \Rightarrow g'_{\mu\nu}=\lambda^{2}g_{\mu\nu}$.
This leads to $g^{\mu\nu}$ having the co-tensor power of $n=-2$ in order to have the 
Kronecker $\delta$ as scale invariant object ($g_{\mu\nu}g^{\nu\rho}=\delta_{\mu}^{\rho}$).
That is, a co-tensor is of power $n$ when, upon local scale change, it satisfies:
$l'\rightarrow\lambda(x)l :\;  Y'_{\mu\nu}\rightarrow\lambda^{n}Y_{\mu\nu}$,
That is a scale-invariant EGR quantity denoted by primed quantity can be obtained from 
a WIG co-tensor of power $n$  upon its multiplication by the $\lambda^{n}$ factor.

In the Dirac co-calculus, this results in the appearance of the ``connexion'' vector field $\kappa_{\mu}$ 
in the  covariant derivatives of scalars, vectors, and tensors (see Table \ref{Table1});
where the usual Christoffel symbol $\Gamma_{\mu\alpha}^{\nu}$ is replaced by
\begin{equation}
^{*}\Gamma_{\mu\alpha}^{\nu}=\Gamma_{\mu\alpha}^{\nu}
+g_{\mu\alpha}k^{\nu}-g_{\mu}^{\nu}\kappa_{\alpha}-g_{\alpha}^{\nu}\kappa_{\mu}.
\label{eq:Christoffel*}
\end{equation}

The corresponding equation of the geodesics within the WIG 
was first introduced in 1973 by \citet{Dirac73} 
and in the weak-field limit of Weyl gauge change redivided in 1979 by \citet{MBouvier79}
($u^{\mu}=dx^{\mu}/{ds}$ is the four-velocity):
\begin{equation}
u_{*\nu}^{\mu}=0\Rightarrow\frac{du^{\mu}}{ds}
+^{*}\varGamma_{\nu\rho}^{\mu}u^{\nu}u^{\rho}
+\kappa_{\nu}u^{\nu}u^{\mu}=0\,.
\label{eq:geodesics*}
\end{equation}
This geodesic equation has also been derived from 
reparametrization-invariant action in 1978 by \citet{BouvM78} given by 
$\delta\mathcal{A}=\intop_{P_{0}}^{P_{1}}\delta\left(d\widetilde{s}\right)
=\int\delta\left(\beta ds\right)=\int\delta\left(\beta\frac{ds}{d\tau}\right)d\tau=0.$

\begin{table}[h]
\caption{\label{Table1} 
Derivatives for co-tensors of power $n$ defined via $Y'_{\mu\nu}\rightarrow\lambda^{n}Y_{\mu\nu}$ when $l'\rightarrow\lambda(x)l$.}
\begin{ruledtabular}
\begin{tabular}{ll}
\textbf{Co-Tensor Type} & \textbf{Mathematical Expression} \\
\hline
co-scalar & $S_{*\mu}=\partial_{\mu}S-n\kappa_{\mu}S$,\\
co-vector & $A_{\nu*\mu}=\partial_{\mu}A_{\nu}-\;^{*}\Gamma_{\nu\mu}^{\alpha}A_{\alpha}-n\kappa_{\nu}A_{\mu}$,\\
co-covector & $A_{*\mu}^{\nu}=\partial_{\mu}A^{\nu}+\;^{*}\Gamma_{\mu\alpha}^{\nu}A^{\alpha}-nk^{\nu}A_{\mu}$.\\
\end{tabular}
\end{ruledtabular}
\end{table}

\subsection{Consequences of Going beyond the EGR}
Before we go into a specific examples, 
such as FLRW cosmology and weak-field limit, we would like to make few remarks.
By using \eqref{eq:Christoffel*} in \eqref{eq:geodesics*}, one can see that 
the usual EGR equations of motion receive extra terms proportional to the four-velocity 
and \mbox{its normalization}:

\begin{equation}
\frac{du^{\mu}}{ds}+\varGamma_{\nu\rho}^{\mu}u^{\nu}u^{\rho}
=(\kappa\cdot u)u^{\mu}-(u\cdot u)\kappa^{\mu}
\label{eq:geodesics+}
\end{equation}

In the weak-field approximation within the SIV, one assumes an isotropic and homogeneous space
for the derivation of the terms beyond the usual Newtonian equations \citep{BouvM78}. 
As seen from \eqref{eq:geodesics+}, the result is a velocity dependent extra term  $\kappa_0\vec{v}$ with 
$\kappa_0=-\dot{\lambda}/\lambda$ and $\vec\kappa=0$ 
due to the assumption of isotropic and homogeneous space. 
At this point, it is important to stress that the usual normalization 
for the four-velocity, $u\cdot u=\pm1$ with sign 
related to the signature of the metric tensor $g_{\mu\nu}$,
is a special choice of $s$-parametrization---the proper-time
parametrization $\tau$.

Recently, similar $\kappa_0\vec{v}$ term was derived as a consequence of 
non-reparametrization invariant mathematical modeling 
but without the need for a weak-field approximation. 
The effect is due to un-proper time parametrization manifested 
as velocity dependent fictitious acceleration \citep{sym13030379}.
In this respect, the term $\kappa_0 \vec{v}$ is necessary for the restoration of 
the broken symmetry - the re-parametrization invariance of a process under study. 
To demonstrate this, one can apply an arbitrary time re-parametrization 
$\lambda= dt/d\tau$; then, the first term on the LHS of \eqref{eq:geodesics+} becomes:

\begin{equation}
\lambda \frac {d }{dt} \left(\lambda \frac {d\vec{r}}{dt} \right)\, 
= \lambda^2 \frac {d^2\vec{r}}{dt^2}\,+ \lambda\dot{\lambda}\frac{d\vec{r}}{dt} .
\label{eq:re-parametrization}
\end{equation}

By moving the term linear in the velocity to the RHS, 
dividing by $\lambda^2$, and by using $\kappa(t)=- \dot{\lambda}/\lambda$, 
one obtains a $\kappa_0 \vec{v}$-like term on the RHS.
If we were to do such manipulation in the absence of $\kappa_0 \vec{v}$ on the RHS of \eqref{eq:geodesics+}, 
then the term will be generated, while if $\tilde\kappa$ was present then it will be transformed
$\tilde\kappa\rightarrow \kappa+\tilde\kappa$.

Furthermore, unlike in  SIV,  where one can justify  $\lambda(t)=t_0/t$ \cite{Maeder17a}, 
for re-parametrization symmetry the time dependence of $\lambda(t)$ could be arbitrary.
Finally, as discussed in \cite{sym13030379}, the extra term $\kappa_0 \vec{v}$ is
not expected to be present when the time parametrization of the process is the 
proper time of the system. Thus, a term of the form $\kappa \vec{v}$ can be viewed 
as restoration of the re-parametrization symmetry and 
an indication of un-proper time parametrization of a process under consideration.

\subsection{Scale Invariant Cosmology}
The scale invariant cosmology equations were first introduced in 1973 by \citet{Dirac73}
and then re-derived in 1977 by \citet{Canuto77}. The equations are based on the
corresponding expressions of the Ricci tensor and the relevant extension of the Einstein equations.
The conformal transformation ($g'_{\mu\nu}=\lambda^{2}g_{\mu\nu}$)  
of the metric tensor $g_{\mu\nu}$ in the more general Weyl's framework into
Einstein's framework,  where the metric tensor is $g'_{\mu\nu}$, induces a simple relation between 
the Ricci tensor and scalar in the Weyl's Integrable Geometry and the Einstein GR   framework  
(using prime to denote Einstein GR   framework  objects):
\[
R_{\mu\nu}=R'_{\mu\nu}-\kappa_{\mu;\nu}-\kappa_{\nu;\mu}-2\kappa_{\mu}\kappa_{\nu}
+2g_{\mu\nu}\kappa^{\alpha}\kappa_{\alpha}-g_{\mu\nu}\kappa_{;\alpha}^{\alpha}\;
\qquad\text{and}\qquad
R=R'+6\kappa^{\alpha}\kappa_{\alpha}-6\kappa_{;\alpha}^{\alpha}\,.
\]
By considering the Einstein equation $R_{\mu\nu}-\frac{1}{2}\ g_{\mu\nu}R=-8\pi GT_{\mu\nu}-\Lambda\,g_{\mu\nu}$
along with the above expressions, one gets:
\begin{equation}
R'_{\mu\nu}-\frac{1}{2}\ g_{\mu\nu}R'-\kappa_{\mu;\nu}-\kappa_{\nu;\mu}
-2\kappa_{\mu}\kappa_{\nu}+2g_{\mu\nu}\kappa_{;\alpha}^{\alpha}-g_{\mu\nu}\kappa^{\alpha}\kappa_{\alpha}=
-8\pi GT_{\mu\nu}-\Lambda\,g_{\mu\nu}\,.
\label{field}
\end{equation}

The relationship $\Lambda=\lambda^{2}\Lambda_{\mathrm{E}}$ of $\Lambda$ in WIG
to the Einstein cosmological constant $\Lambda_{\mathrm{E}}$ in the EGR
was present in the original form of the equations to provide explicit scale invariance.
This relationship makes explicit the appearance of $\Lambda_{\mathrm{E}}$ as invariant scalar (in-scalar), 
as then one has $\Lambda\,g_{\mu\nu}=\lambda^{2}\Lambda_{\mathrm{E}}\,g_{\mu\nu}=\Lambda_{\mathrm{E}}\,g'_{\mu\nu}$.

The above equation \eqref{field} ia a generalization of the original Einstein GR equation.
Thus, they have an even larger class of local gauge symmetries that need to be fixed by a gauge choice.
In Dirac's work, the gauge choice was based on the large numbers hypothesis.
Here, we will discuss a different gauge choice - the SIV gauge.

The corresponding scale-invariant FLRW based cosmology equations within 
the WIG   framework  were first introduced in 1977 by \citet{Canuto77}:
\begin{equation}
\frac{8\,\pi G\varrho}{3}=\frac{k}{a^{2}}+\frac{\dot{a}^{2}}{a^{2}}+2\,\frac{\dot{\lambda}\,\dot{a}}{\lambda\,a}
+\frac{\dot{\lambda}^{2}}{\lambda^{2}}-\frac{\Lambda_{\mathrm{E}}\lambda^{2}}{3}\,,\quad\text{and}\quad
-8\,\pi Gp=\frac{k}{a^{2}}+2\frac{\ddot{a}}{a}+2\frac{\ddot{\lambda}}{\lambda}+\frac{\dot{a}^{2}}{a^{2}}
+4\frac{\dot{a}\,\dot{\lambda}}{a\,\lambda}-\frac{\dot{\lambda^{2}}}{\lambda^{2}}-\Lambda_{\mathrm{E}}\,\lambda^{2}\,.\label{E2p}
\end{equation}

These equations clearly reproduce the standard FLRW equations in the limit $\lambda=const=1$.
The scaling of $\Lambda$ was recently used to revisit the 
Cosmological Constant Problem within quantum cosmology \cite{GueorM20}.
The conclusion of \cite{GueorM20} is that our universe is unusually large, 
given that the expected mean size of all universes, 
where Einstein GR holds, is expected to be of a Plank scale.
In the study, $\lambda=const$ was a key assumption as the 
universes were expected to obey the Einstein GR equations.
What the expected mean size of all universes would be if the condition  $\lambda=const$ is relaxed, as for a WIG-universes ensemble, remains an open question.

\subsection{The Scale Invariant Vacuum Gauge ($T=0$ and $R'=0$)} 

The idea of the Scale Invariant Vacuum was introduced first in 2017 by \citet{Maeder17a}.
It is based on the fact that, for Ricci flat ($R'_{\mu\nu}=0$) Einstein GR vacuum ($T_{\mu\nu}=0$),
one obtains from \eqref{field} the following equation for the vacuum:
\begin{equation}
\kappa_{\mu;\nu}+\kappa_{\nu;\mu}+2\kappa_{\mu}\kappa_{\nu}
-2g_{\mu\nu}\kappa_{;\alpha}^{\alpha}+g_{\mu\nu}\kappa^{\alpha}\kappa_{\alpha}=\Lambda\,g_{\mu\nu}.
\label{SIV}
\end{equation}

For homogeneous and isotropic WIG-space $\partial_{i}\lambda=0$;
therefore, only $\kappa_{0}=-\dot{\lambda}/\lambda$ and its time
derivative $\dot{\kappa}_{0}=-\kappa_{0}^{2}$ can be non-zero.
As a corollary of \eqref{SIV}, one can derive the following set of equations \cite{Maeder17a}:
\begin{eqnarray}
\ 3\,\frac{\dot{\lambda}^{2}}{\lambda^{2}}\,=\Lambda\,,\;
\mathrm{and}\quad2\frac{\ddot{\lambda}}{\lambda}-\frac{\dot{\lambda}^{2}}{\lambda^{2}}\,=\Lambda\,,\quad
\mathrm{or}\quad\frac{\ddot{\lambda}}{\lambda}\,=\,2\,\frac{\dot{\lambda}^{2}}{\lambda^{2}}\,,\;
\mathrm{and}\quad\frac{\ddot{\lambda}}{\lambda}-\frac{\dot{\lambda}^{2}}{\lambda^{2}}\,=\frac{\Lambda}{3}\,.\label{SIVeqs}
\end{eqnarray}

One could derive these equations by using the time and space components of the equations \eqref{SIV} or 
by looking at the relevant trace invariant along with the relationship $\dot{\kappa}_{0}=-\kappa_{0}^{2}$.
Any pair of these equations is  sufficient to prove the other pair of equations. 

\begin{thm}
Using any one pair of two SIV Equations \eqref{SIVeqs} along with $\Lambda=\lambda^{2}\Lambda_{E}$ one has: 
\begin{equation}
\Lambda_{E}=3\frac{\dot{\lambda^{2}}}{\lambda^{4}},\quad\mathrm{with}\quad\frac{d\Lambda_{E}}{dt}=0.
\label{SIV-gauge}
\end{equation}
\end{thm}
 
\begin{cor}
The solution of the SIV equations is:
$\lambda=t_{0}/t,$
with $t_0=\sqrt{3/\Lambda_{E}}$ and $c=1$ for the speed of light.
\end{cor}

Upon the use of the SIV gauge, first in  2017 by \citet{Maeder17a}, one observes that 
{\it the cosmological constant disappears} from Equations \eqref{E2p}:  
\begin{equation}
\frac{8\,\pi G\varrho}{3}=\frac{k}{a^{2}}+\frac{\dot{a}^{2}}{a^{2}}+2\,\frac{\dot{a}\dot{\lambda}}{a\lambda}\,,\quad\text{and}\quad
-8\,\pi Gp=\frac{k}{a^{2}}+2\frac{\ddot{a}}{a}+\frac{\dot{a^{2}}}{a^{2}}+4\frac{\dot{a}\dot{\lambda}}{a\lambda}\,.\label{E2}
\end{equation}

\section{Comparisons and Applications}
The predictions and outcomes of the SIV paradigm were confronted 
with observations in a series of papers by the current authors.
Highlighting the main results and outcomes is the subject of current section.

\subsection{Comparing the Scale Factor $a(t)$ within  $\Lambda$CDM and SIV}
\begin{figure}[h]
\includegraphics[width=0.8\textwidth]{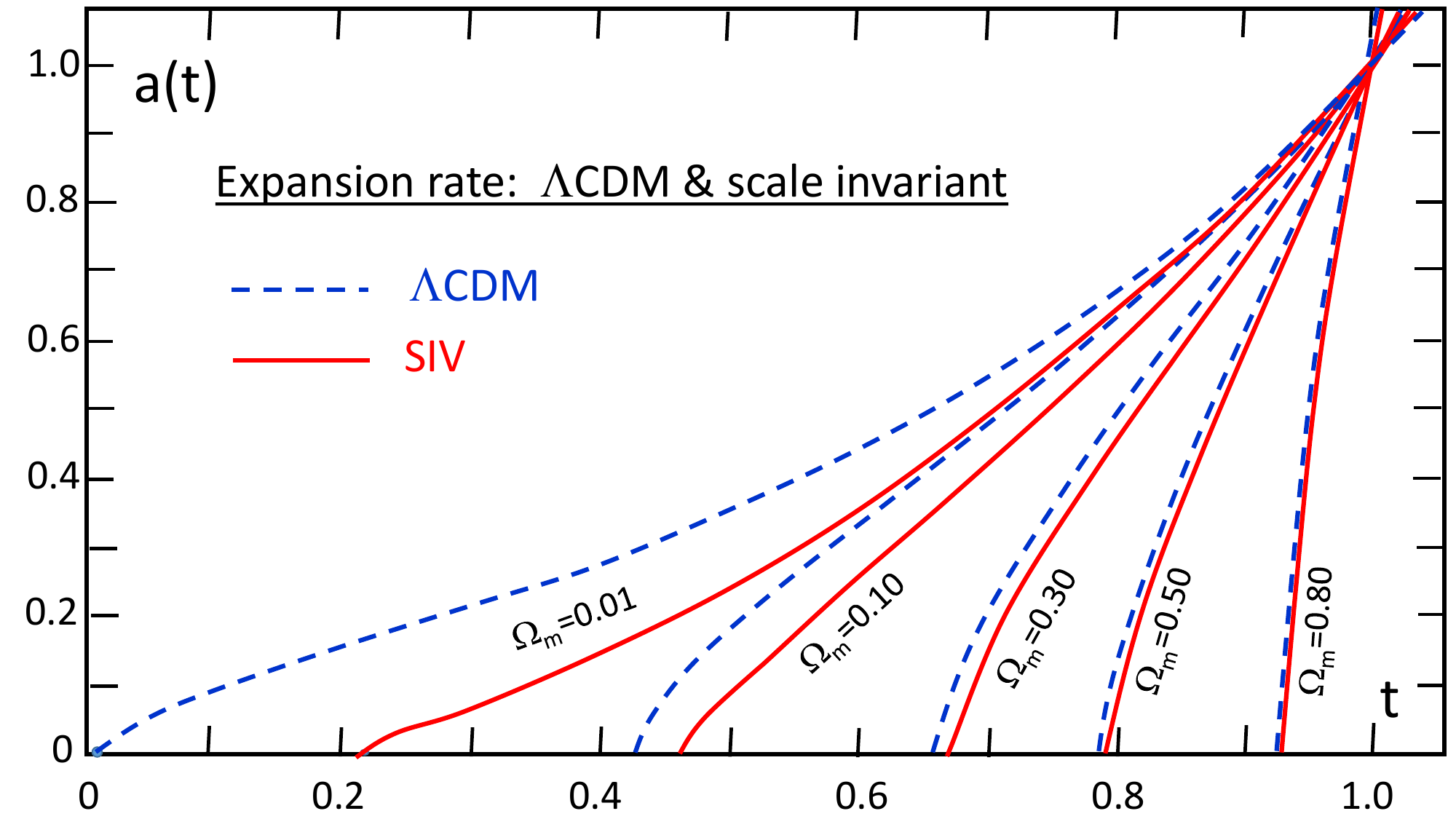} 
\caption{\label{rates} 
Expansion rates $a(t)$ as a function of time $t$ in the flat
($k=0$) $\Lambda$CDM and SIV models in the matter dominated era.
The curves are labeled by the values of $\Omega_{\mathrm{m}}$.}
\end{figure}

Upon arriving at the SIV cosmology Equations \eqref{E2}, 
along with the gauge fixing \eqref{SIV-gauge},  which implies $\lambda=t_0/t$ 
with $t_0$ indicating the current age of the universe since the Big-Bang ($a=0$ at $t_\text{in}<t_0$), 
the implications for cosmology were first discussed \mbox{by \citet{Maeder17a}} and 
later reviewed by \citet{MaedGueor20a}. The most important point in 
comparing $\Lambda$CDM and SIV cosmology models is the existence of 
SIV cosmology with slightly different parameters but almost the same 
curve for the standard scale parameter $a(t)$ 
when the time scale is set so that  $t_0=1$ now 
\cite{Maeder17a,MaedGueor20a}. 
As seen in Figure~\ref{rates}, the difference between the 
$\Lambda$CDM and SIV models declines for increasing matter densities.
Furthermore, for any $\Lambda$CDM curve at some $\Omega'_m$ there is a matching 
SIV curve at some $\Omega_m<\Omega'_m$. Thus, SIV needs less total matter to 
produce the same scale-factor evolution.

\subsection{Application to Scale-Invariant Dynamics of Galaxies}

The next important application of the scale-invariance at cosmic scales is
the derivation of a universal expression for the
Radial Acceleration Relation (RAR) of $g_{\mathrm{obs}}$  and $g_{\mathrm{bar}}$.
That is, the relation between the observed gravitational acceleration $g_{\mathrm{obs}}=v^2/r$
and the baryonic matter acceleration due to the 
standard Newtonian gravity $g_\mathrm{N}$ by \cite{MaedGueor20b}:
\begin{equation}
g\,=\,g_{\mathrm{N}}+\frac{k^{2}}{2}+\frac{1}{2}\sqrt{4g_{\mathrm{N}}k^{2}+k^{4}}\,,
\label{sol}
\end{equation}
where $g=g_{\mathrm{obs}}$, $g_N=g_{\mathrm{bar}}$.
For $g_{\mathrm{N}} \gg k^{2} : g \rightarrow g_{\mathrm{N}}$
but for $g_{\mathrm{N}}\rightarrow 0 \Rightarrow g \rightarrow k^{2}$ is a constant.

\begin{figure}[h]
\includegraphics[width=0.7\textwidth]{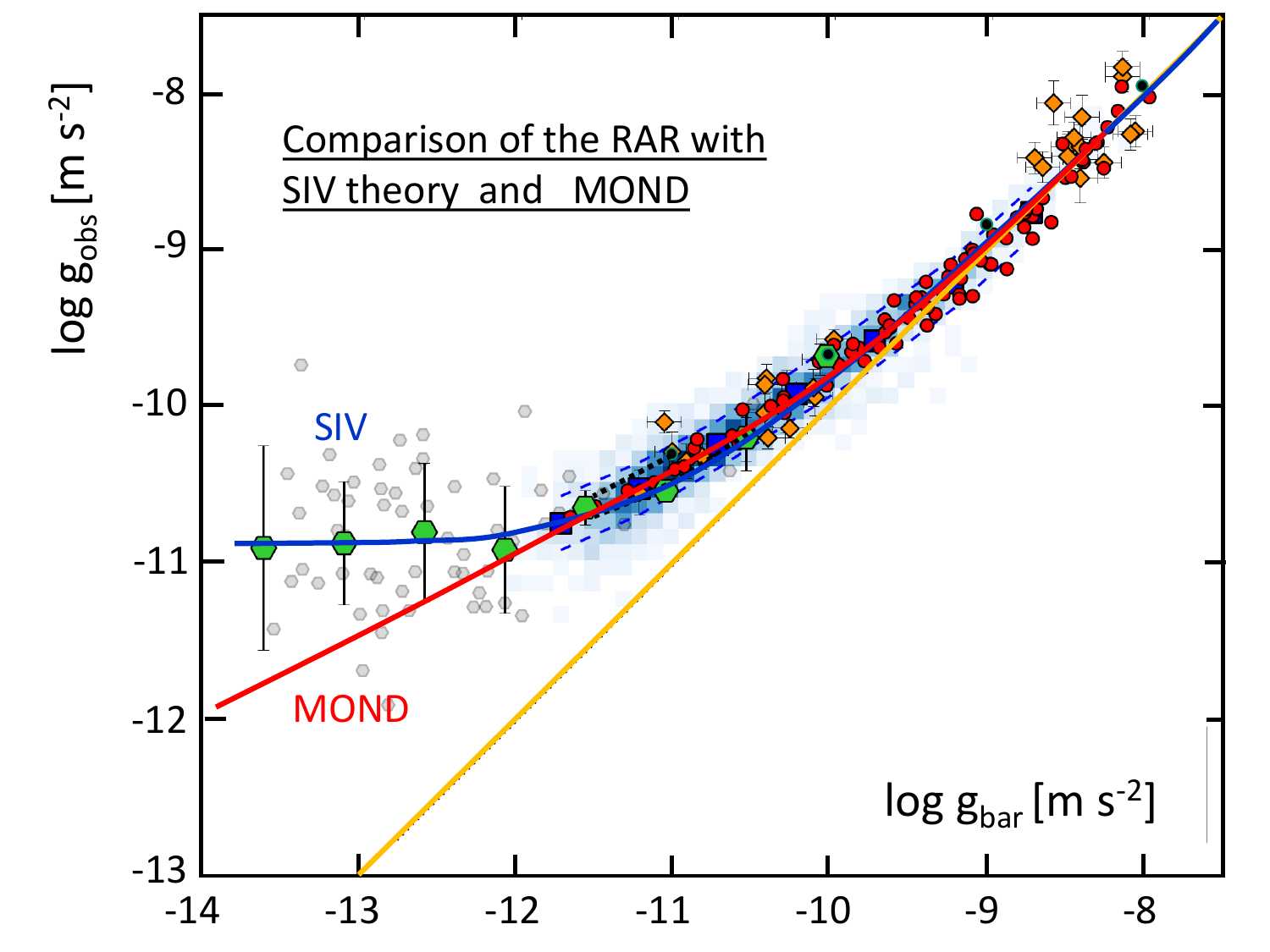} 
\caption{
Radial Acceleration Relation (RAR) for the galaxies studied by Lelli et al. (2017).
Dwarf Spheroidals  as binned data (big green hexagons), 
along with MOND (red curve), and SIV (blue curve) model predictions.
The orange curve shows the 1:1-line for $g_{\mathrm{obs}}$  and $g_{\mathrm{bar}}$.
Due to to the smallness of  $g_{\mathrm{obs}}$  and $g_{\mathrm{bar}}$ 
the application of the $\log$ function results in negative numbers; 
thus, the corresponding axes' values are all negative.}
\label{gobs} 
\end{figure}

As seen in Figure~\ref{gobs}, MOND deviates significantly for the data on the Dwarf Spheroidals.
This is well-known problem in MOND due to the need of two different interpolating functions,
one in galaxies and one at cosmic scales. The SIV expression \eqref{sol} resolves 
this issue via one universal parameter $k^2$ related to the gravity at large distances 
\cite{MaedGueor20b}. Even more, one can actually show that 
MOND is a peculiar case of the SIV theory \citep{Maeder23}.

The expression \eqref{sol} follows from the Weak Field Approximation (WFA) of the SIV upon
utilization of the Dirac co-calculus in the derivation of the geodesic equation within the 
relevant WIG \eqref{eq:geodesics*} (see \citet{MaedGueor20b} for more details, 
as well as the original derivation in \citet{MBouvier79}):  
\begin{equation}
g_{ii}=-1,\;g_{00}=1+2\Phi/c^{2}\Rightarrow\varGamma_{00}^{i}
=\frac{1}{2}\frac{\partial g_{00}}{\partial x^{i}}=\frac{1}{c^{2}}\frac{\partial\Phi}{\partial x^{i}}\;\Rightarrow\;
\frac{d^{2}\overrightarrow{r}}{dt^{2}}=-\frac{GM}{r^{2}}\frac{\overrightarrow{r}}{r}
+\kappa_{0}(t)\frac{d\overrightarrow{r}}{dt}.\label{Nvec}
\end{equation}
where $i\in{1,2,3}$, while the potential $\Phi=GM/r$ is scale invariant.

By considering the scale-invariant ratio of the correction term $\kappa_0(t) \, \vec{v}\, $ 
to the usual Newtonian term in \eqref{Nvec}, one has:
$x=\frac{\kappa_0 v r^{2}}{GM}=\frac{H_{0}}{\xi}\frac{v \,r^{2}}{GM}
= \, \frac{H_0}{\xi}  \frac{(r \, g_{\mathrm{obs}})^{1/2}}{g_{\mathrm{bar}}}
\sim\frac{g_{\mathrm{obs}}-g_{\mathrm{bar}}}{g_{\mathrm{bar}}}\,.$
Then, by utilizing an explicit scale invariance for canceling the proportionality factor:
${\left(\frac{g_{\mathrm{obs}}-g_{\mathrm{bar}}}{g_{\mathrm{bar}}}\right)_{2}}\div
{\left(\frac{g_{\mathrm{obs}}-g_{\mathrm{bar}}}{g_{\mathrm{bar}}}\right)_{1}}\,=
\,\left(\frac{g_{\mathrm{obs, 2}}}{ g_{\mathrm{obs, 1}}}\right)^{1/2}\,
\left(\frac{g_{\mathrm{bar,1}}}{g_{\mathrm{bar,2}}}\right)\,,$
by setting $g=g_{\mathrm{obs, 2}}$, $g_N=g_{\mathrm{bar,2}}$, 
and by collecting all the system-1 terms in $k=k_{(1)}$, 
then one arrives at \eqref{sol} by solving for $g$ in
$\frac{g}{g_{\mathrm{N}}}-1=k_{(1)}\frac{g^{1/2}}{g_{\mathrm{N}}}$
and keeping the bigger root (the positive sign in $\pm\sqrt{\;\dots\;}$ factor).

\subsection{Growth of the Density Fluctuations within the SIV}

Another interesting result was the possibility of a fast growth of the density fluctuations 
within the SIV \citep{MaedGueor19}. This study accordingly modifies  
the relevant equations such as the continuity equation, Poisson equation, and Euler equation 
within the SIV framework. Here, we outline the main equations and the relevant results.

By using the notation ${ \kappa}=\kappa_0=-\dot{\lambda}/\lambda=1/t$, 
the corresponding Continuity, Poisson, and Euler equations are:
\[
\frac{\partial \rho}{\partial t}+ \vec{\nabla}\cdot (\rho \vec{v}) 
=  \kappa  \left [\rho+ \vec{r} \cdot \vec{\nabla} \rho \right]\,,\; 
\vec{\nabla}^{2}\Phi=\triangle\Phi=4\pi G \varrho \label{eq:Continuity+Poisson},\;
\frac{d\vec{v}}{dt}=\frac{\partial\vec{v}}{\partial t}+\left(\vec{v}\cdot\vec{\nabla}\right)\vec{v}=
-\vec{\nabla}\Phi-\frac{1}{\rho}\vec{\nabla}p+ \kappa \vec{v}\, .
\]

For a density perturbation $\varrho(\vec{x},t)\, = \, \varrho_{b}(t)(1+\delta(\vec{x},t))$
the above equations result in:
\begin{eqnarray}
\dot{\delta} + \vec{\nabla} \cdot \dot{\vec{x}} =\kappa \vec{x} \cdot \vec{\nabla} \delta = n \kappa(t) \delta,&&
\vec{\nabla}^{2}\Psi = 4\pi Ga^{2}\varrho_{b}\delta,\;\label{D1}\quad
\ddot{\vec{x}}+ 2 H \dot{\vec{x}}+ (\dot{\vec{x}}\cdot \vec{\nabla}) \dot{\vec{x}} 
= -\frac{\vec{\nabla} \Psi}{a^2} +\kappa(t) \dot{\vec{x}}.
\end{eqnarray}

\begin{figure}[h]
\includegraphics[width=0.6\textwidth]{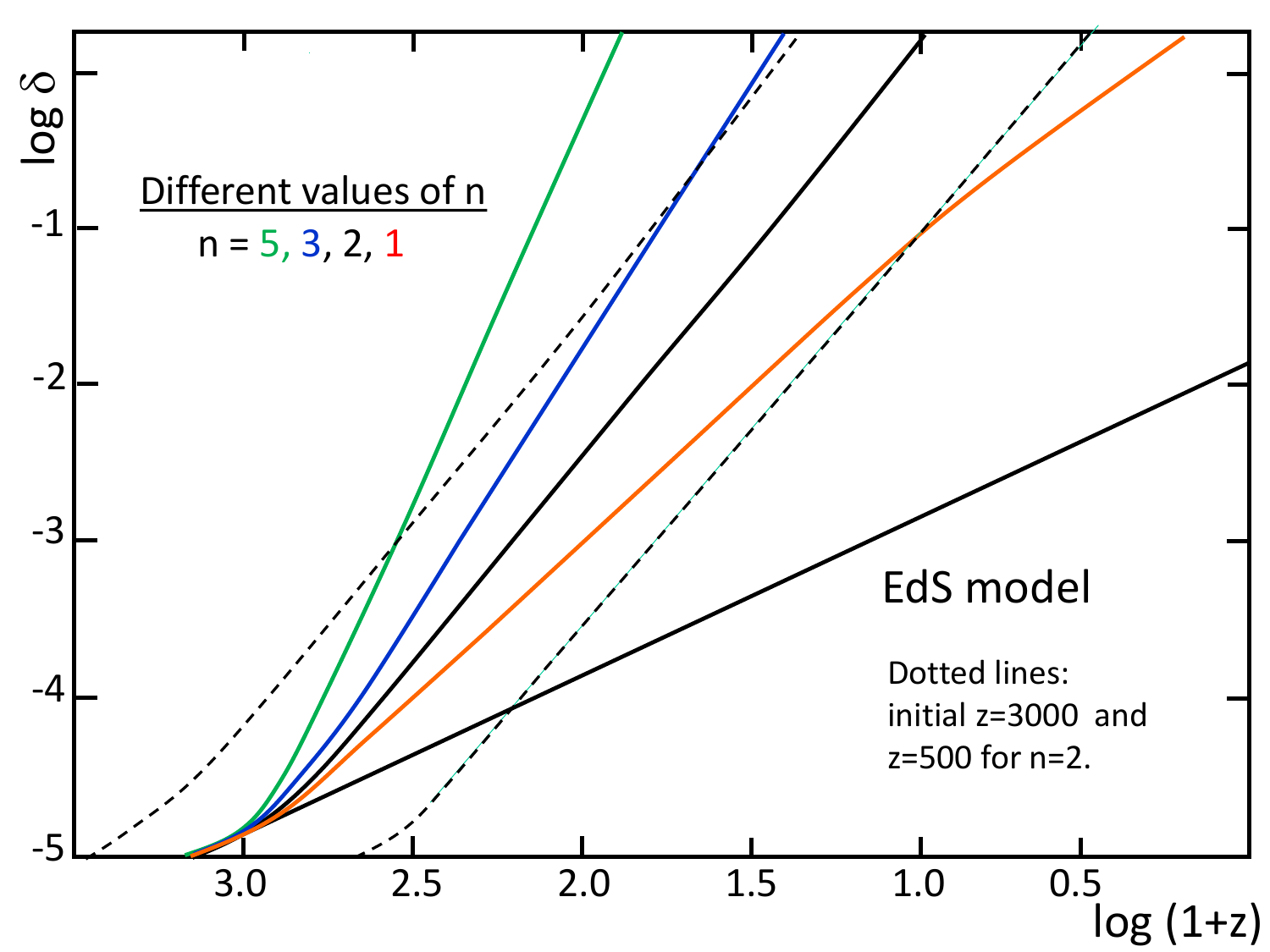} 
\caption{
The growth of density fluctuations for different values of parameter
$n$ (the gradient of the density distribution in the nascent cluster),
for an initial value $\delta=10^{-5}$ at $z=1376$ and $\Omega_{\mathrm{m}}=0.10$.
The initial slopes are those of the EdS models. The two light broken
curves show models with initial $(z+1)=3000$ and 500, with same $\Omega_{\mathrm{m}}=0.10$
and $n=2$. These dashed lines are to be compared to the black continuous
line of the $n=2$ model. All the three lines for $n=2$ are very
similar and \mbox{nearly parallel}.
Due to to the smallness of  $\delta$  the application of the $\log$ function results in negative numbers; 
resulting in negative vertical axis.}
\label{variousn} 
\end{figure}

The final result $\ddot{\delta}+ (2H -(1+n)  \kappa  )\dot{\delta} 
= 4\pi G \varrho_{b}\delta + 2 n  \kappa  (H- \kappa ) \delta$
recovers the standard equation 
in the limit of $  \kappa  \rightarrow 0$.
The simplifying assumption $\vec{x} \cdot \vec{\nabla} \delta(x) = n \delta(x)$ 
in \eqref{D1} introduces the parameter $n$ that measures the perturbation
type (shape). For example, a spherically symmetric perturbation would have $n=2$.
As seen in Figure~\ref{variousn}, perturbations for various values of $n$ are 
resulting in faster growth of the density fluctuations within the SIV than in  
the Einstein--de Sitter model, even at relatively law matter densities. 
Furthermore, the overall slope is independent of the choice of recombination epoch $z_\mathrm{rec}$.
The behavior for different $\Omega_m$  is also very interesting, and is shown and discussed in detail
by \citet{MaedGueor19}. 

\subsection{Big-Bang Nucleosynthesis within the Scale Invariant Vacuum Paradigm}

The SIV paradigm has been recently applied to the Big-Bang Nucleosynthesis 
using the known analytic expressions for the expansion factor $a$ and
the plasma temperature $T$ as functions of the SIV time $\tau$ since the Big-Bang
when $a(\tau=0)=0$ \cite{VG&AM'23}. The results have been compared to the standard
BBNS as calculated via the PRIMAT code \cite{PRIMAT}. Potential SIV-guided deviations from the 
local statistical equilibrium were also explored in ref. \cite{VG&AM'23}. 
Overall, it was found that smaller than usual baryon 
and non-zero dark matter content, by a factor of three to five times reduction, 
result in compatible to the standard light elements abundances (Table \ref{Table2}).

The SIV analytic expressions for $a(T)$ and $\tau(T)$ were utilized to study the BBNS within the SIV paradigm \cite{VG&AM'23, Maeder18}.
The functional behavior is very similar to the standard model within PRIMAT except during the very early universe 
where electron-positron annihilation and neutrino processes affect the $a(T)$ function (see Table I and Fig. 2 in ref. \cite{VG&AM'23}). 
The distortion due to these effects encoded in the function $S(T)$ could be incorporated by considering the SIV paradigm 
as a background state of the universe where these processes could take place.
It has been demonstrated that incorporation of the  $S(T)$ 
within the SIV paradigm results in a compatible outcome with the standard BBNS
see the last two columns of Table  \ref{Table2}; 
furthermore, if one is to fit the observational data the result is $\lambda\approx1$
for the SIV parameter $\lambda$ (see last column of Table \ref{Table2} with $\lambda=\text{FRF}\approx1$). 
However, a pure SIV treatment (the middle three columns) results in $\Omega_b\approx1\%$ and less total matter,
either around $\Omega_m\approx23\%$  when all the $\lambda$-scaling connections are utilized (see Table  \ref{Table2} column 6), 
or around $\Omega_m\approx6\%$ without any $\lambda$-scaling factors 
(see column 5 of Table  \ref{Table2}).
The need to have $\lambda$ close to 1 is not an indicator of dark matter content but
indicates the goodness of the standard PRIMAT results that allows only for $\lambda$ close to 1 
as an augmentation, as such this leads to a light but important improvement in D/H 
as seen when comparing columns three with eight and nine.

\begin{table}[h]
\begin{tabular}{|c|cc|ccc|ccc|}
\hline\hline 
Element & Obs. & PRMT & $a_{SIV}$ & fit & fit* & $\bar{a}/\lambda$&  fit* & fit \\
\hline 
 \text{H} & 0.755 & 0.753 & 0.805 & 0.755 & 0.849 & 0.75 & 0.753 & 0.755 \\
 $Y_P=4Y_{\text{He}}$ & 0.245 & 0.247 & 0.195 & 0.245 & 0.151 & 0.25 & 0.247 & 0.245 \\
 $\text{D/H}\times10^5$ & 2.53 & 2.43 & 0.743 & 2.52 & 2.52 & 1.49 & 2.52 & 2.53 \\
\hline  
 $^3\text{He/H}\times10^5$ & 1.1 & 1.04 & 0.745 & 1.05 & 0.825 & 0.884 & 1.05 & 1.04 \\
 $^7\text{Li/H}\times10^{10}$ & 1.58 & 5.56 & 11.9 & 5.24 & 6.97 & 9.65 & 5.31 & 5.42 \\
\hline  
$N_{\text{eff}}$ & 3.01 & 3.01 & 3.01 & 3.01 & 3.01 & 3.01 & 3.01 & 3.01 \\
$\eta_{10}$ & 6.09 & 6.14 & 6.14 & 1.99 & 0.77 & 1.99 & 5.57 & 5.56 \\
 FRF  & 1 & 1 & 1 & 1 & 1.63 & 1 & 1 & 1.02 \\
 \text{m\v{T}} & 1 & 1 & 1 & 1 & 0.78 & 1 & 1 & 0.99 \\
 \text{Q/\v{T}} & 1 & 1 & 1 & 1 & 1.28 & 1 & 1 & 1.01 \\
$\Omega _b$\; [\%] & 4.9 & 4.9 & 4.9 & 1.6 & 0.6 & 1.6 & 4.4 & 4.4 \\
 $\Omega _m$\;[\%]  & 31 & 31 & 31 & 5.9 & 23 & 5.9 & 86 & 95 \\
 $\sqrt{\chi _{\epsilon}^2}$ & N/A & 6.84 & 34.9 & 6.11 & 14.8 & 21.9 & 6.2 & 6.4 \\
\hline\hline 
\end{tabular}
\caption{\label{Table2}
The observational uncertainties are  1.6\% for $Y_{P}$, 1.2\% for D/H, 18\% for T/H, and 19\% for Li/H.
FRF is the forwards rescale factor for all reactions, 
while \text{m\v{T}}  and \text{Q/\v{T}} are the corresponding rescale factors
in the revers reaction formula based on the local thermodynamical equilibrium. 
The SIV $\lambda$-dependences are used when these factors are different from 1;
that is, in the sixth and ninth columns where FRF=$\lambda$, 
\text{m\v{T}}= $\lambda^{-1/2}$, and \text{Q/\v{T}}= $\lambda^{+1/2}$. 
The columns denoted by fit contain the results for perfect fit 
on $\Omega _b$ and $\Omega _m$  to $^{4}$He and D/H,
while fit* is the best possible fit on $\Omega _b$ and $\Omega _m$ 
to the $^{4}$He and D/H observations 
for the model considered as indicated in the columns four and seven.
The last three columns are usual PRIMAT runs with modified $a(T)$ 
such that $\bar{a}/\lambda=a_{SIV}/{S}^{1/3}$,
where $\bar{a}$ is the PRIMAT's $a(T)$ for the decoupled neutrinos case.
Column seven is actually $a_{SIV}/{S}^{1/3}$,
but it is denoted by $\bar{a}/\lambda$ to remind us 
about the relationship $a'=a\lambda$;
the run is based on $\Omega _b$ and $\Omega _m$ from column five.
The smaller values of $\eta_{10}$ are due to smaller 
$h^2\Omega_b$, as seen by noticing that 
$\eta_{10}/\Omega_b$ is always $\approx 1.25$.}
\end{table}

The SIV paradigm suggests specific modifications to the reaction rates, 
as well as the functional temperature dependences of these rates,
that need to be implemented to have consistence between the 
Einstein GR frame and the WIG (SIV) frame.
In particular, the non-in-scalar factor $T^\beta$ in the reverse reactions rates 
may be affected the most due to the SIV effects. 
As shown in \cite{VG&AM'23}, the specific dependences studied,   
within the assumptions made within the SIV model, resulted in three times less baryon matter,
usually around $\Omega_b\approx1.6\%$  and less total matter - around $\Omega_m\approx6\%$.
The lower baryon matter content leads to also a lower photon to baryon ratio $\eta_{10}\approx2$
within the SIV, which is three tines less that the standard value of $\eta_{10}=6.09$.
The overall results indicated insensitivity to the specific 
$\lambda$-scaling dependence of the \text{m\v{T}}-factor in the reverse reaction expressions within $T^\beta$ terms. 
Thus, one may have to explore further the SIV-guided $\lambda$-scaling relations as done
for the last column in  Table  \ref{Table2}, however, this would require the 
utilization of the numerical methods used by PRIMAT and as such will take us away from the 
SIV-analytic expressions explored that provided a simple model for understanding the BBNS within the SIV paradigm. 
Furthermore, it will take us further away from the accepted local statistical equilibrium and
may require the application of the reparametrization paradigm that seems to  result in SIV like 
equations but does not impose a specific form for $\lambda$ \citep{sym13030379}.

\subsection{SIV and the Inflation of the Early Universe}

The latest published result within the SIV paradigm is the presence of inflation stage at the very early universe
$t\approx 0$ with a natural exit from inflation in a later time $t_\mathrm{exit}$ with 
value related to the parameters of the inflationary potential  \citep{SIV-Inflation'21}.
The main steps towards these results are outlined below.

If we go back to the first of the general scale-invariant cosmology Equations \eqref{E2p},
we can identify a vacuum energy density expression that relates 
the Einstein cosmological constant with the energy density
as expressed in terms of $\kappa=-\dot{\lambda}/\lambda$ by using the SIV result 
\eqref{SIV-gauge}. The corresponding vacuum energy density $\rho$, with $C=3/(4\pi G)$, is then:
\[
\rho=\frac{\Lambda}{8\pi G}=\lambda^{2}\rho'=\lambda^{2}\frac{\Lambda_{E}}{8\pi G}
=\frac{3}{8\pi G}\frac{\dot{\lambda}^{2}}{\lambda^{2}}
=\frac{C}{2}\dot{\psi}^{2}\,.
\]

This provides a natural connection to inflation within the SIV 
via $\dot{\psi}=-\dot{\lambda}/\lambda$ or $\psi\propto\ln(t)$. 
The equations for the energy density, pressure, and Weinberg's condition for inflation 
within the standard inflation \citep{Guth81,Linde95,Linde05,Weinberg08} are:
\begin{eqnarray}
\left.\begin{array}{c}\rho\\ p \end{array}\right\} 
=\frac{1}{2}\dot{\varphi}^{2}\pm V(\varphi),\label{rp}\;
\mid\dot{H}_{\mathrm{infl}}\mid\,\ll H_{\mathrm{infl}}^{2}\,.
\label{cond}
\end{eqnarray}

\noindent
If we make the identification between the standard inflation above with the 
fields within the SIV (using $C=3/(4\pi G)$):
\begin{eqnarray}
\dot{\psi}=-\dot{\lambda}/\lambda, & \varphi\leftrightarrow\sqrt{C}\,\psi, \quad
V\leftrightarrow CU(\psi), & U(\psi)\,=\,g\,e^{\mu\,\psi}\,.
\label{SIV-identification}
\end{eqnarray}

\noindent
Here, $U(\psi)$ is the inflation potential with strength $g$ and field ``coupling'' $\mu$. 
One can evaluate the Weinberg's condition for inflation \eqref{cond} 
within the SIV framework \citep{SIV-Inflation'21}, and the result is:
\begin{equation}
\frac{\mid\dot{H}_{\mathrm{infl}}\mid}{H_{\mathrm{infl}}^{2}}\,
=\,\frac{3\,(\mu+1)}{g\,(\mu+2)}\,t^{-\mu-2}\ll1\,{\normalcolor for}\ \mu<-2,\ {\normalcolor and}\ t\ll t_{0}=1.
\label{crit}
\end{equation}
\noindent
From this expression, one can see that there is a graceful exit from inflation at the \mbox{later time:}
\begin{equation}
t_\mathrm{exit}\approx\sqrt[n]{\frac{n\ g}{3(n+1)}}\qquad \mathrm{with} \qquad n=-\mu-2>0,
\label{t_exit}
\end{equation}
when the Weinberg's condition for inflation \eqref{cond} is not satisfied anymore.
For more details, we refer the reader to the derivation of the equation \eqref{crit} 
presented in our previous papers \citep{SIVProgress22,SIV-Inflation'21}.


\section{Conclusions and Outlook}
From the highlighted results in the previous section on various comparisons and potential applications, 
we see that the \textit{SIV cosmology is a viable alternative to $\Lambda$CDM.} 
In particular, within the SIV gauge (\ref{E2}) \textit{the cosmological constant disappears}.  
There are diminishing differences in the values of the scale factor $a(t)$ 
within $\Lambda$CDM and SIV at higher densities as emphasized in the 
discussion of (Figire~\ref{rates}) \cite{Maeder17a,MaedGueor20a}. 
Furthermore, the SIV also shows consistency for $H_0$ and the age of the universe,  
and the m-z diagram is well satisfied---see \citet{MaedGueor20a} for details.

Furthermore, \textit{the SIV provides the correct RAR for dwarf spheroidals} (Figure~\ref{gobs}) 
while MOND is failing, and dark matter cannot account for the phenomenon \cite{MaedGueor20b}.
Therefore, it seems that  \textit{within the SIV, dark matter is not needed to seed the growth of structure} 
in the universe, as there is a fast enough growth of the density fluctuations as seen in (Figure~\ref{variousn}) 
and discussed in more detail by \citet{MaedGueor19}.

As to the BBNS within the SIV, our main conclusion is that the SIV paradigm provides a concurrent model of the BBNS
that is compatible to the description of $^4$He, D/H, T/H, and $^7$Li/H achieved  in the standard BBNS.
It suffers of the same $^7$Li  problem as in the standard BBNS but also suggests a possible 
SIV-guided departure from local statistical equilibrium which could be a fruitful direction to be explored 
towards the resolution of the $^7$Li  problem.

In our study on the inflation within the SIV cosmology \cite{SIV-Inflation'21},
we have identified a connection of the scale factor $\lambda$, and its rate of change, 
with the inflation field $\psi \rightarrow \varphi\,,\; \dot{\psi}=-\dot{\lambda}/\lambda$ \eqref{SIV-identification}.
As seen from \eqref{crit}, \textit{inflation of the very-very early universe ($t\approx0$) 
is natural, and SIV predicts a graceful exit from inflation} (see \eqref{t_exit})!

Some of the obvious future research directions are related to the primordial nucleosynthesis,
where preliminary results show a satisfactory comparison between SIV and observations \cite{VG&AM'23,Maeder18}. 
The recent success of the R-MOND in the description of the CMB \cite{2021PhRvL.127p1302S}, 
after the initial hope and concerns \cite{Skordis2006}, is very stimulating and demands testing SIV cosmology 
against the MOND and $\Lambda$CDM successes in the description of the CMB,  
the Baryonic Acoustic Oscillations, etc.

Another important direction is the need to understand the physical meaning and interpretation of the 
conformal factor $\lambda$. As we pointed out in the Motivation Section, a general conformal factor $\lambda(x)$
seems to be linked to Jordan–Brans–Dicke scalar-tensor theory that leads to a varying Newton's constant G,
which has not been found in nature. Furthermore, a spacial dependence of  $\lambda(x)$ opens the door
to local field excitations that should manifest as some type of fundamental scalar particles. 
The Higgs boson is such a particle, but a connection to Jordan–Brans–Dicke theory seems
a far fetched idea. On the other hand, the assumption of isotropy and homogeneity of space forces 
$\lambda(t)$ to depend only on time, which is not in any sense similar to the usual fundamental fields 
we are familiar with.

In this respect, other less obvious research directions are related to the exploration of SIV within the solar system due to the 
high-accuracy data available, or exploring further and in more detail the possible connection of SIV 
with the re-parametrization invariance. For example, it is already known by \citet{sym13030379} 
that un-proper time parametrization can lead to a SIV-like equation of motion \eqref{eq:geodesics+}
and the relevant weak-field version \eqref{Nvec}. 

\begin{acknowledgments}
A.M. expresses his gratitude to his wife for her patience and support. 
V.G. is extremely grateful to his wife and daughters for their understanding and family support  during the various stages of the research presented. 
This research did not receive any specific grant from funding agencies in the public, commercial, or not-for-profit sectors.
V.G. is grateful to the organizers of the PRIT'23 conference for their kind accommodations needed for V.G. to be able to present online this research results.
\end{acknowledgments}

\end{document}